\documentclass[iop]{emulateapj}
\usepackage{natbib}
\usepackage{mathrsfs}
\usepackage{url}
\newcommand{\IUE}{{\it IUE}}
\newcommand{\HST}{{\it HST}}
\newcommand{\kms}{\ifmmode {\rm km~s}^{-1} \else km~s$^{-1}$\fi}
\newcommand{\ergs}{\ifmmode {\rm erg~ s}^{-1} \else erg~s$^{-1}$\fi}
\newcommand{\ergscm}{\ifmmode {\rm erg~s}^{-1} \else erg~s$^{-1}$ cm$^{-2}$\fi}
\newcommand{\Msun}{\ifmmode {\rm M}_{\odot} \else M$_{\odot}$\fi }
\newcommand{\Lsun}{\ifmmode {\rm L}_{\odot} \else L$_{\odot}$\fi}
\newcommand{\qo}{\ifmmode q_{\rm o} \else $q_{\rm o}$\fi}
\newcommand{\Ho}{\ifmmode H_{\rm o} \else $H_{\rm o}$\fi}
\newcommand{\ho}{\ifmmode h_{\rm o} \else $h_{\rm o}$\fi}

\newcommand{\vFWHM}{\ifmmode v_{\mbox{\tiny FWHM}} \else
                    $v_{\mbox{\tiny FWHM}}$\fi}
\newcommand{\CCF}{\ifmmode F_{\it CCF} \else $F_{\it CCF}$\fi}
\newcommand{\ACF}{\ifmmode F_{\it ACF} \else $F_{\it ACF}$\fi}
\newcommand{\Halpha}{\ifmmode {\rm H}\alpha \else H$\alpha$\fi}
\newcommand{\Hbeta}{\ifmmode {\rm H}\beta \else H$\beta$\fi}
\newcommand{\Hgamma}{\ifmmode {\rm H}\gamma \else H$\gamma$\fi}
\newcommand{\Hdelta}{\ifmmode {\rm H}\delta \else H$\delta$\fi}
\newcommand{\Lya}{\ifmmode {\rm Ly}\alpha \else Ly$\alpha$\fi}
\newcommand{\Lyb}{\ifmmode {\rm Ly}\beta \else Ly$\beta$\fi}
\newcommand{\HeI}{\ifmmode {\rm He}\,{\sc i}\,\lambda5876 \else
	          He\,{\sc i}\,$\lambda5876$\fi}
\newcommand{\HeII}{\ifmmode {\rm He}\,{\sc ii}\,\lambda4686 \else 
	           He\,{\sc ii}\,$\lambda4686$\fi}
\newcommand{\hi}{\ifmmode \makebox{{\rm H\,}{\sc i}} \else H\,{\sc i}\fi}

\newcommand{\heii}{\ifmmode \makebox{{\rm He}\,{\sc ii}} \else He\,{\sc ii}\fi}

\newcommand{\ciii}{\ifmmode {\rm C}\,{\sc iii} \else C\,{\sc iii}\fi}

\def\fake2{\hphantom{3}}

\shorttitle{AGN STORM IV. Anomalous Line Behavior in NGC\,5548}
\shortauthors{Goad et al.}

\begin{document}

\title{Space Telescope and Optical Reverberation Mapping Project.\\
IV.\ Anomalous Behavior of the Broad Ultraviolet Emission Lines in NGC\,5548}

\author {M.~R.~Goad\altaffilmark{1}, K.~T.~Korista\altaffilmark{2},
  G.~De~Rosa\altaffilmark{3,4,5}, G.~A.~Kriss\altaffilmark{5,6},
  R.~Edelson\altaffilmark{7}, A.~J.~Barth\altaffilmark{8},
  G.~J.~Ferland\altaffilmark{9}, C.~S.~Kochanek\altaffilmark{3,4},
  H.~Netzer\altaffilmark{10}, B.~M.~Peterson\altaffilmark{3,4}
% ...
}
\altaffiltext{1}{University of Leicester, Department of Physics and
  Astronomy, Leicester, LE1 7RH, UK} 
\altaffiltext{2}{Department of
  Physics, Western Michigan University, 1120 Everett Tower, Kalamazoo,
  MI 49008-5252, USA} 
\altaffiltext{3}{Department of Astronomy, The
  Ohio State University, 140 W 18th Ave, Columbus, OH 43210, USA}
\altaffiltext{4}{Center for Cosmology and AstroParticle Physics, The
  Ohio State University, 191 West Woodruff Ave, Columbus, OH 43210,
  USA} 
\altaffiltext{5}{Space Telescope Science Institute, 3700 San
  Martin Drive, Baltimore, MD 21218, USA} 
\altaffiltext{6}{Department
  of Physics and Astronomy, The Johns Hopkins University, Baltimore,
  MD 21218, USA} 
\altaffiltext{7}{Department of Astronomy, University
  of Maryland, College Park, MD 20742-2421, USA}
\altaffiltext{8}{Department of Physics and Astronomy, 4129 Frederick
  Reines Hall, University of California, Irvine, CA 92697, USA}
\altaffiltext{9}{Department of Physics and Astronomy, The University
  of Kentucky, Lexington, KY 40506, USA} 
\altaffiltext{10}{School of
  Physics and Astronomy, Raymond and Beverly Sackler Faculty of Exact
  Sciences, Tel Aviv University, Tel Aviv 69978, Israel}
% ...

\footnotetext[13]{NSF Postdoctoral Research Fellow}
\footnotetext[21]{Einstein Fellow}
\footnotetext[64]{Packard Fellow}
\begin{abstract}
During an intensive {\it Hubble Space Telescope (HST)} Cosmic Origins
Spectrograph (COS) UV monitoring campaign of the Seyfert~1 galaxy
NGC\,5548 performed from 2014 February to July, the normally highly
correlated far-UV continuum and broad emission-line variations
decorrelated for $\sim60$--70 days, starting $\sim75$ days after the
first \HST/COS observation. Following this anomalous state, the flux
and variability of the broad emission lines returned to a more normal
state.  This transient behavior, characterised by significant deficits
in flux and equivalent width of the strong broad UV emission lines, is
the first of its kind to be unambiguously identified in an active
galactic nucleus reverberation mapping campaign.  The largest
corresponding emission-line flux deficits occurred for the
high-ionization collisionally excited lines, C~{\sc iv} and Si~{\sc
  iv}($+$O~{\sc iv]}), and also He~{\sc ii}($+$O~{\sc iii]}), while
    the anomaly in Ly$\alpha$ was substantially smaller. This pattern
    of behavior indicates a depletion in the flux of photons with
    $E_{\rm ph} > 54$~eV, relative to those near 13.6~eV.  We suggest
    two plausible mechanisms for the observed behavior: (i) temporary
    obscuration of the ionizing continuum incident upon BLR clouds by
    a moving veil of material lying between the inner accretion disk
    and inner BLR, perhaps resulting from an episodic ejection of
    material from the disk, or (ii) a temporary change in the
    intrinsic ionizing continuum spectral energy distribution
    resulting in a deficit of ionizing photons with energies $>$
    54~eV, possibly due to a transient restructuring of the
    Comptonizing atmosphere above the disk. Current evidence appears
    to favor the latter explanation.
 \end{abstract}

\keywords{galaxies: active --- galaxies: individual (NGC\,5548) ---
galaxies: nuclei --- galaxies: Seyfert }


\section{INTRODUCTION}
\label{section:intro}

One of the most secure correlations in studies of active galactic
nuclei (AGN) is that found between the ultraviolet (UV)--optical
continuum variations and the UV--optical broad emission-line (BEL)
variations.  This simple relation is causal in nature.  The
UV--optical broad emission lines arise in high-density, fast-moving
gas that is photoionized by an intense source of ionizing continuum
radiation originating from a disk of material feeding the central
supermassive black hole. Intrinsic variations in the incident ionizing
continuum flux translate into correlated changes in the BEL strengths
and their ratios. That an AGN's observed UV/optical continuum
variations closely track the largely unobservable driving continuum is
demonstrated by the observed positive correlations between the flux
variations in the emission lines and the observed UV/optical
continuum. Positive correlations are also observed between the
UV/optical and X-ray variations (Clavel et al. 1992; Marshall et
al. 1997; Peterson et al. 2000; Uttley et al. 2003; Edelson et
al. 2015, hereafter Paper~{\sc ii}).

This ubiquitous property has proven extremely useful because the
correlated continuum--emission-line variations may be used to probe
the spatial distribution and kinematics of the BEL
gas, probing size scales (roughly a few microarcseconds) inaccessible
via more conventional means.  This technique, commonly referred to as
reverberation mapping (RM; see Blandford \& McKee 1982), in its
simplest form can reveal the characteristic ``size'' of the broad
emission-line region (BLR) for lines of very different ionization
state and arising in gas with a wide range in density, thereby
mapping out the density and ionization structure of the BLR. However,
its real power manifests when the measured delays between the
continuum and emission-line variations are coupled with velocity
information. Time-resolved spectroscopy can reveal not only the bulk
motion of the line-emitting gas, but has been usefully exploited to
measure the mass of the central black hole in $\sim60$ AGN (e.g.,
Denney et~al.\ 2010; Bentz et~al.\ 2010b; Grier et al.\ 2012; Kaspi et
al.\ 2000, 2007; Peterson et~al.\ 2002; Peterson 2014, and references
therein; see also Bentz \& Katz 2015 for a comprehensive list of RM
sources).

RM is now a mature field.  Notable improvements have been made in
target selection, campaign design, and observing strategies.  Coupled
with new techniques developed to isolate the variable broad emission
lines from contaminating components and major advances in methods for
recovering the emission-line response function, these have facilitated
improved dynamical mass estimates for several AGN (Barth et al.\ 2011;
Bentz et al.\ 2014; Brewer et~al. 2011; Denney et al.\ 2006, 2009a,
2010; Pancoast et al.\ 2012, 2014a,b; Zu et al. 2011, 2013a,b), as
well as black hole mass determinations for possibly super-Eddington
sources (Grier et al. 2012; Du et al. 2014, 2015; Hu et al. 2015; Wang
et al. 2014), objects with purported slim disks and exhibiting a
different mode of accretion.  Significantly, RM campaigns are now
producing velocity-delay maps of sufficient fidelity that the gas
spatial configuration and dynamics can be discerned (Denney
et~al.\ 2009b; Bentz et~al.\ 2010a,b; Skielboe et~al.\ 2015).


In 2014, the AGN Space Telescope and Optical Reverberation Mapping (STORM)
collaboration undertook the most intensive RM
experiment to date. Awarded 179 orbits in Cycle 22 with
\HST/COS\,, the aim was to provide the first high-fidelity
velocity-resolved delay maps for the strong, broad UV emission lines in
an AGN. The chosen target, NGC\,5548, has a well-documented history of
correlated large-amplitude continuum and BEL
variations at both UV and optical wavelengths (Clavel et al.\ 1991;
Korista et~al.\ 1995; Peterson et al.\ 2002, and references
therein). The \HST/COS observations are presented by De Rosa et al.\ (2015,
hereafter Paper~{\sc i}) and are of an unprecedented quality. These
data were supported by photometric observations with {\it
  Swift} (Paper~{\sc ii}), as well as by
ground-based photometric (Fausnaugh et al. 2016,
hereafter Paper~{\sc iii}) and spectroscopic (Pei et al., in
prep.; hereafter Paper~{\sc v}) observations.

This remarkable dataset has revealed a number of interesting and
unexpected results. Chief among these is a temporary
breakdown in one of the fundamental tenets of RM. Approximately 75
days after the start of the \HST/COS\, campaign, the continuum and
emission-line variations appeared to decorrelate. This transient
phenomenon, characterized by a significant and anomalous depression in
the BEL flux and a notable reduction in
emission-line variability lasting $\sim65$--70~days, is the subject
of this investigation.

\section{Reverberation mapping}

The primary goal of RM is to use correlated continuum and BEL 
variations to solve the transfer equation,

\begin{equation}
L(V,t) = \int_{0}^{\infty} \Psi(V,\tau)C(t-\tau)d\tau \, ,
\end{equation}

\noindent where $\Psi(V,\tau)$, the transfer function (or convolution
kernel), maps the {\em driving\/} continuum light curve at earlier
times $C(t-\tau)$ onto the observed emission-line light curve $L(V,t)$
at the current epoch.

To eliminate nuisance background components (e.g., the host-galaxy
contribution to the UV/optical continuum bands and contaminating
narrow emission-line contributions), this equation has traditionally
been solved in its linearized form

\begin{equation}
\Delta L(V,t) = \int_{0}^{\infty} \Psi^{\prime}(V,\tau)\Delta
C(t-\tau) d\tau \, ,
\end{equation}

\noindent where $\Delta L(V,t)$ and $\Delta C(t-\tau)$ respectively 
represent the variable component of $L(V,t)$ and $C(t-\tau)$ relative
to some average background value. Here, the quantity of interest is
$\Psi^{\prime}(V,\tau)$, the emission-line ``response function.'' In
general, $\Psi^{\prime}(V,\tau) \neq \Psi(V,\tau)$.

Underpinning this simple relation, and a prerequisite for a
successful RM campaign, is that the continuum and emission-line
variations are causally related. More specifically the basic
assumptions of RM are as follows.

\begin{enumerate}
\item{At the distance of the BLR, the driving continuum source appears
  point-like.}

\item{The ionizing continuum and emission-line photons propagate
  freely at the speed of light --- i.e., the BLR filling factor is
  small.}

\item{There exists a simple causal (though not necessarily linear)
  relation between the observed continuum and emission-line
  variations.}

\item{The observed continuum variations, modulo some scale factor and
  temporal smoothing function, are a suitable proxy for the
  unobserved driving ionizing continuum variations.}

\item{The dominant timescale is the light-crossing time.}

\end{enumerate}

\noindent Violation of any of these assumptions will impact our
ability to recover $\Psi^{\prime}(V,\tau)$.  Fortunately, in {\em the
  vast majority of previous RM experiments\/}, these assumptions are
broadly consistent with the data.

\section{Following the energy}
\subsection{Anomalous Broad Emission-Line Variations}

The strongest evidence to date for the violation of one of the above
assumptions (specifically, assumption 4) appeared during the
middle of the AGN STORM campaign, upon comparison of the UV continuum
and BEL light curves (Paper~{\sc i})\footnote{The only
  anomalies reported to date have been attributed solely to changes in
  the continuum reprocessing efficiency of BLR clouds
  (e.g., Maoz 1992; Sparke 1993; Grier et al. 2008). In these cases,
  while the line responsivity changed, the emission-line and continuum
  variations remain significantly correlated.}.  While the UV
continuum and BEL variations were well-correlated at
the start and end of the \HST\, observing campaign, they appeared to
decorrelate starting approximately midway through the campaign. It
was for this reason that in Paper~{\sc i} a separate analysis was
performed on the data for the two halves of the campaign.  We
illustrate this in Figure~1c, where we compare a modified
version of the C~{\sc iv} BEL light curve (colored
points) with the continuum light curve at 1157\,\AA. We use
emission-line light curves based on the {\it direct integrations\/}
over the bulk of the broad emission line reported in Paper~{\sc i},
and modified as described below.

For C~{\sc iv} and all other emission lines, we have {\em shifted the
  entire emission-line light curve\/} according to the measured delay
between the emission line and the 1157\,\AA\ continuum band as
determined for {\em the first 75~days\/} of the \HST\, campaign. We
assume that this delay provides an adequate description of the delay
between the continuum and emission-line variations for the latter half
of the campaign.  For the measured delays we use the peak and centroid
of the cross-correlation function (hereafter CCF, see Table~1). The
CCF centroids are calculated over the range in delay for which the CCF
coefficient exceeds 80\% of the peak value. Uncertainties are estimated using
the model-independent FR/RSS Monte Carlo method (Peterson et
al.\ 1998), computing 1000 realizations of each light curve, assuming
random sampling and full replacement.
Next, we remove an estimate for contaminating narrow emission-line
flux (Table~1, column 4; based on measurements of low flux-state
spectra of NGC\,5548 first reported by Crenshaw et al.\, 1993; see also
Goad \& Koratkar 1998), and then rescale the emission-line light curve
to the mean continuum flux over the first 75 days of the \HST\,
campaign. Finally, we scale the emission-line variability amplitude by
dividing through by the measured responsivity for each line (\S3.3 and
Table~2, column 2).  In so doing, the continuum and emission-line
variations should then be of similar amplitude (they will be identical
if the response function were simply a delta function shifted in
time).

Figures~1a--1c indicate that starting from day 75 (cyan
points), the C~{\sc iv} flux relative to our proxy measure of the
ionizing continuum flux drops significantly and then remains
anomalously low for $\sim40$ days (red points), before recovering
(magenta points) and again showing normal, correlated behavior (green
points).  Additionally, during the anomalous period (red points), the
emission-line variability amplitude, relative to the observed
continuum variations, is also significantly suppressed.  Similar
behavior, though of varying strength, is seen {\em for all of the
  strong, broad UV emission lines\/} (\S3.4), and is indicative of a
dramatic decline in the continuum reprocessing efficiency of the 
line-emitting gas, as inferred by the observed continuum at 1157\,\AA.  We
have examined the 1989 \IUE/SWP,LWP\, (Clavel et al. 1991) and 1993 \HST/FOS\,
 (Korista et al. 1995) monitoring campaigns on NGC~5548 to
look for similar effects and found none. To our knowledge, the AGN
STORM campaign is the first for which a decorrelation between the
observed UV continuum and UV BEL variations of such
long duration has been reported.

\subsection{BEL Responsivity and Reprocessing Efficiency}

Significant changes in the continuum reprocessing efficiency of an
emission line are best revealed through studying the emission-line
responsivity and/or equivalent width (EW). The time-averaged
emission-line responsivity $\eta_{\rm eff}$ is the power-law index
relating the observed continuum, preferably the band closest in
wavelength to the ionizing continuum, and BEL fluxes,
$F_{\rm cont}$ and $F_{\rm line}$:

\begin{equation}
F_{\rm line} \propto F_{\rm cont}^{\eta_{\rm eff}} \, .
\end{equation}

\noindent The exponent ${\eta_{\rm eff}}$ is normally measured after
first removing contaminating contributions from nonvariable
background components (e.g., the narrow emission lines and starlight
from the host galaxy), and after correcting for the mean delay between
the continuum and emission-line variations (Pogge \& Peterson 1992;
Gilbert \& Peterson 2003; Goad, Korista \& Knigge 2004), in order to
minimize the effects of geometrical dilution.
\noindent Alternatively, the emission-line EW and 
the continuum may be related by

\begin{equation}
{\rm EW}_{\rm line} \propto F_{\rm cont}^{\beta}  \, ,
\end{equation}

\noindent where $\beta=\eta_{\rm eff}-1$.  The relationship between
line responsivity and emission-line EW is thus made clear.  For a
strictly linear (1:1) response, $\eta_{\rm eff} = 1$, $\beta=0$, and
the emission-line EW remains constant relative to the continuum
flux variations. If $\eta_{\rm eff}=0$, the emission line does not
respond to variations in the driving continuum. In general, $\eta_{\rm
  eff} < 1$ for most emission lines (i.e., a nonlinear response), and
hence $\beta < 0$. Thus, the majority of emission lines will show an
intrinsic Baldwin effect (Kinney et~al. 1990). The measured value of
$\eta_{\rm eff}$ will also depend upon the reference continuum band
(since the amplitudes of the continuum variations are larger at shorter
wavelengths --- e.g., Wamsteker et al. 1990; Clavel et al. 1991;
Papers~{\sc ii} \& {\sc iii}) and the degree of geometrical dilution
(Gilbert \& Peterson 2003; Goad \& Korista 2014, 2015).

\subsection{Measuring $\eta_{\rm eff}$ and Identifying the Anomaly}

For the continuum light curve against which the emission-line delays
and their responsivities will be measured, we use the \HST\, continuum-band 
measurements centered at 1157\,\AA\ (Paper~{\sc iii}), as this is
the wavelength band accessible with \HST/COS\, that lies nearest to
the driving ionizing continuum and has negligible host-galaxy
contamination\footnote{Relative to the driving ionizing continuum
  variations, the observed UV and optical continuum variations are of
  lower amplitude and smeared in time (e.g., Papers~{\sc ii} \& {\sc
    iii}), and may at some wavelengths be significantly contaminated
  by more slowly varying background components.}.  For illustrative
purposes, we focus on the strong, broad C~{\sc iv} emission line,
but we treat all other emission lines in the same fashion.  To measure
$\eta_{\rm eff}$ we follow the procedure of Goad, Korista, \& Knigge
(2004). Using the delay-corrected BEL light curve we
reconstruct the continuum flux associated with the BEL
flux at the current epoch by interpolation, adopting a weighted average
of the two bracketing continuum points. The appropriate weights are
derived from the first-order structure function of the continuum light
curve (Kawaguchi et al. 1998; Paltani 1999; Goad, Korista, \& Knigge
2004). Continuum data associated with emission-line data shifted to
epochs before the start of the campaign are determined by a linear
extrapolation. Emission-line points beyond the end of the campaign can
be determined in a similar fashion. We do not use these extrapolated
points (either emission line or continuum) when measuring the
time-averaged BEL responsivity $\eta_{\rm eff}$, preferring
instead to exclude these data from the analysis. Uncertainties in the
reconstructed continuum and emission-line points were determined from
a structure-function analysis of their respective light curves (for
details, see Goad, Korista, \& Knigge 2004).

Having corrected the data for the mean delay $\langle \tau \rangle$,
we then fit the relation

\begin{equation}
\log F_{\rm line} =  A  + \eta_{\rm eff} [ \log F_{\rm cont} (1157\mathrm{\AA}) ] \, 
\end{equation}

\noindent to points lying outside of the anomaly (blue and green
points only), assuming uncertainties in both the ordinate and abscissa. For
C~{\sc iv}, the best-fit slope gives $\eta_{\rm eff} = 0.25\pm0.01$
(Figure~1a, solid red line). In Table~2, column 2, we report the
best-fit slopes for the time-averaged responsivity $\eta_{\rm eff}$
and their $1\sigma$ uncertainties for all of the broad emission lines
reported in Paper~{\sc i}. Quoted values are the centroid and
1$\sigma$ uncertainties determined from bootstrap resampling of the
(corrected) light curves (10,000 realizations) with full replacement.

Figure~1a indicates that during the anomaly (epochs 75--140) there
is a dramatic and uncorrelated reduction in the C~{\sc iv}
emission-line flux and response amplitude to continuum
variations. Note in particular the island of red points well below the
main $\log F_{\rm line}$--$\log F_{\rm cont}$ relation with
$\eta_{\rm eff}\approx 0$, indicating either (i) an absence of
emission-line response to continuum variations, or (ii) that the
ionizing continuum responsible for driving the observed BEL
variations varies less than the 1157\,\AA\ continuum.  Also,
note that the journey away from the main relation (cyan points) into
the anomaly forms a continuous path (as indicated by the cyan
arrows). Curiously, when exiting the anomaly (magenta points), the data
follow a similar path but in the reverse direction (magenta arrows),
before rejoining the normal $F_{\rm line}$--$F_{\rm cont}$
behavior\footnote{An animated GIF of this behavior is available in
  the online material.}. There are indications that the $\log
F_{\rm line}$--$\log F_{\rm cont}$ relation is somewhat steeper at
the end of the campaign (green points) than during the first 75 days
(blue points), though some of this effect may result from the use of a
single lag when correcting for the continuum--emission-line delays.
Larger emission-line responsivities ($\eta_{\rm eff}$) are usually
associated with the emergence from lower ionizing continuum flux
states (Korista \& Goad 2004; Goad \& Korista 2014, 2015).

Figure~1b illustrates the relationship between the
1157\,\AA\ continuum flux and C~{\sc iv} equivalent width,
referenced to the 1157\,\AA\ continuum. The best-fit slope in
this relation (fitting to data on either side of the anomaly; i.e., blue
and green points only) is $\beta = -0.75\pm0.01$, as expected since
$\beta = \eta_{\rm eff} - 1$. Once again, the color-coded behavior
described above is evident. There is an island of red points lying
below the main relation and for which $\beta \approx -1$, while the
cyan and magenta points taken together appear as extensions of the
same relation with a slope intermediate between that of the main
relation and that of the anomaly.  The same general behavior, though
differing in detail and with differing noise levels, is seen
for Ly$\alpha$, Si~{\sc iv}($+$O~{\sc iv}]), and He~{\sc ii}($+$O~{\sc
    iii}]).

\subsection{Measuring the Amplitude of the Anomaly}

In previous AGN RM campaigns, the strong positive correlation observed
between the UV/optical continuum and BEL variations provides
substantive supporting evidence that the UV and optical continuum
variations are reasonable proxies, though smaller in amplitude and
smeared in time, for the variable driving ionizing continuum.  The
strong positive correlation observed between the continuum and
emission-line variations pre- and post-anomaly for this campaign
suggests that over these time periods, the UV/optical continuum
variations are also suitable proxies for the variable driving ionizing
continuum. If this interpretation is correct, these data may then be
used to infer the ``expected'' form of the emission-line light curve
during the anomaly --- that is, as if the anomaly had not occurred.

Using the
$\beta$ values reported in Table~2, column 3, we reconstruct the
emission-line flux corresponding to the observed continuum flux at
1157\,\AA\ and shifted in time according to the delay reported in
Table~1, column 2\footnote{A more thorough treatment would instead
  solve for the transfer function using those epochs bracketing the
  anomaly, and then apply forward modeling to infer the expected
  emission-line fluxes during the anomalous period. This will be
  explored in the future but is beyond the scope of the current
  work.}.  In Figure~1d we display a comparison between the observed
(colored points) and reconstructed (black triangles) emission-line
light curve for C~{\sc iv}. In Figures~2a--2d we show the observed
(colored points) and reconstructed (black triangles) emission-line
light curves for all of the strong UV emission lines reported in
Paper~{\sc i}, here normalized to their mean values over the first 75
days of the campaign. There are indications (e.g., Figures 2a--2d)
that the start times, stop times, and duration of the anomaly differs
among the emission lines. If real, timing differences among the
various lines may provide additional clues about the BLR structure. We
defer investigation of this effect to a future paper.

Using the observed and reconstructed emission-line light curves, we can
measure the fraction of line emission lost $f_{\rm lost}$ during the
period of the anomaly, $f_{\rm lost} = \left [ (f_{\rm rec} - f_{\rm
    obs})/f_{\rm rec} \right ] \times 100\%$ (e.g., Figure~1e),
where $f_{\rm obs}$ and $f_{\rm rec}$ represent the observed and
reconstructed emission-line fluxes, respectively. We find time-averaged
fractional losses of $\sim23$\% (Si~{\sc iv}($+$O~{\sc iv}])),
  $\sim21$\% (He~{\sc ii}($+$O~{\sc iii}])), $\sim18$\% (C~{\sc iv}),
    and $\sim9$\% (Ly$\alpha$) during the time period spanning the
    anomaly. The largest flux deficit is found for Si~{\sc iv} and
    He~{\sc ii}, followed closely by C~{\sc iv}, while Ly$\alpha$
    shows the smallest (by a factor 2) flux deficit
    (Figures~2a--2d).  In Table~3, we report the measured
    time-averaged emission-line EWs over the period spanning the
    anomaly (Figure~1a -- red points) and compare it with the
    ``expected'' EW, here defined as the EW measured from the main
    relation (solid red line in Figure~1b) that corresponds to the
    average continuum level measured over the duration of the
    anomalous period [$F_{\rm cont}(1157\mathrm{\AA})= (4.84\pm 0.37)
    \times 10^{-14}$~erg~s$^{-1}$~cm$^{-2}$~\AA$^{-1}$].  The measured
    deficit in EW is largest for Si~{\sc iv}($+$O~{\sc iv}])
      ($\sim24$\%), followed by He~{\sc ii}($+$O~{\sc iii}])
        ($\sim21$\%), C~{\sc iv} ($\sim19$\%), and then
        Ly$\alpha$ ($\sim11$\%), in good agreement with the
        estimates given above.

The analyses described here and illustrated in Figures 1 and 2 can be
further extended to include the strong optical recombination lines
(Paper~{\sc v}, in prep.), and it may also be applied as a function of
projected line-of-sight velocity for all of the strong, broad UV and
optical emission lines. In particular, it may be possible to see the
effects of the transition into, through, and out of the anomaly in the
velocity-resolved data, thereby constraining the BLR geometry and
dynamics. This will require the highest-fidelity light curves, free
from contaminating narrow absorption and emission lines. These are
typically obtained from spectral decomposition of the emission lines
and is work in progress. Finally, we note that any variability in the
equivalent widths of the weak absorption lines associated with the
outflow is unrelated to the anomaly, as clearly indicated by the fact
that He~{\sc ii}($+$O~{\sc iii}]) is strongly affected by the anomaly
  but has no absorption lines.

\section{Discussion}

The observed behavior of the broad emission lines during the middle of
the \HST/COS campaign is consistent with a temporary yet significant
softening of the ionizing continuum spectral energy distribution (SED)
{\em incident\/} upon the BLR lasting $\sim40$ days (Figure 1, red points).
The cyan and magenta points may represent transition phases (each
lasting $\sim1$--2 weeks) straddling the anomalous period. In the
following, we suggest two plausible scenarios to account for the
anomalous behavior of the broad emission lines.

In Figure 2e, we show the light curve for a powerful diagnostic
BEL flux ratio involving the two strongest emission
lines, C~{\sc iv}/Ly$\alpha$ (e.g., Shields et al. 1995; Netzer 1990). This
temperature-sensitive ratio is a measure of the heating-cooling
balance within the BEL clouds, centered on the average
ionizing photon energy of ~$\sim100$~eV, estimated for the central
source in NGC\,5548 (Medhipour et~al.\ 2015).  This flux ratio is thus
particularly sensitive to the SED of the incident ionizing
continuum\footnote{In the absence of changes in the ionizing SED, variations in
  the delay-corrected emission-line flux ratios are mainly due to
  intrinsic differences between their individual line responsivities, and
  any remaining residual reverberation effects (e.g., geometrical
  dilution).}. During the first $\sim80$ days, the C~{\sc
  iv}/Ly$\alpha$ flux ratio varies (weakly) inversely with the
continuum variability, owing to the somewhat larger responsivity of
Ly$\alpha$ (Table~2, column 2).  Approximately coincident with the
onset of the anomaly, this emission-line flux ratio then exhibits a
gradual decline, and continues to decline (on average) for the
duration of the anomalous period, before eventually rising again as
the source behavior returns to normal.

We also show in Figure~2f the light curve for the BEL 
  flux ratio, He~{\sc ii}($+$O~{\sc
  iii]})/Ly$\alpha$. This flux ratio is sensitive to the ratio of
  photons responsible for ionizing the He$^{+}$ ion ($E_{\rm ph} >
  54.4$~eV) to those responsible for ionizing hydrogen ($E_{\rm ph} >
  13.6$~eV). During the first $\sim 87$~days, this flux ratio
  correlates directly with the observed continuum variations, owing to the
  substantially larger responsivity in He~{\sc ii} when compared to
  Ly$\alpha$. It then quickly drops and
  remains approximately constant throughout the anomalous period. This
  behavior is indicative of a reduction in the incident flux of
  photons capable of ionizing He$^{+}$ relative to the flux of
  photons responsible for ionizing hydrogen over the time period of
  anomalous emission-line behavior. We note that the time-dependent
  behavior of this pair of BEL flux ratios is also
  reflected in the comparison of panels a, b, c of Figure~2.

Contemporaneous observations in the soft (0.3--0.8~keV) and hard
(0.8--10~keV) X-ray bands with {\it Swift\/}/XRT (Figure 1 in 
Paper~{\sc ii}) indicate a gradual decline in the soft and hard X-ray
photon fluxes starting around day 75 (the onset of the anomaly),
beyond which the measured flux in the observed UV continuum bands is
trending upward. In Figure~2g, we show the hardness ratio, HR =
(H$-$S)/(H$+$S), constructed from the soft (S) and hard (H) X-ray
bands as defined in Paper~{\sc ii} (see also Medhipour et
al. 2016, in press). Figure~2g indicates that on entering the
anomalous period, the X-ray HR initially increases,
thereafter remaining approximately constant with an average value
similar to that measured at the start and end of the campaign. That
is, there appears to be nothing unusual about the HR
during the anomalous period. In general, the HR shows a marginal trend
downward over the course of the campaign interrupted by two
significant drops (indicative of a softer spectrum), starting
$\sim 50$ days prior to the onset of the anomaly and lasting a
combined total of $\sim 60$ days before returning to the more general
trend.  It is unclear whether these large excursions in HR prior to
the anomaly are in any way related to the anomaly.  Significantly, we
see no color-dependent changes in the UV--optical continuum band
during the time of the anomaly.

On exiting the anomaly, the soft and hard X-ray photon fluxes
appear to trend upward, coincident with the increasing trend in the
C~{\sc iv}/Ly$\alpha$ flux ratio. 
While some of the observed X-ray variability is likely
caused by changes in the absorption of X-ray photons in out-flowing
material lying along our line of sight (Kaastra et al. 2015), these
data are also consistent with a gradual softening of the X-ray--UV
continuum, notably in our line of sight.

All of these observed temporal behaviors are consistent with a temporary
yet significant softening of the incident ionizing continuum both
along our line of sight and toward the BLR.  The larger emission-line
flux and EW deficits exhibited by the collisionally excited lines
(C~{\sc iv} and Si~{\sc iv}($+$O~{\sc iv]})) are consistent with a
  drop in electron temperature that accompanied the softening of the
  ionizing SED.  Together with the larger deficit in the broad He~{\sc
    ii} emission line compared to Ly$\alpha$, the evidence points to
  most of the softening occurring for $E_{\rm ph} >$\, 54~eV.

We suggest two mechanisms to explain the temporary softening of the
incident ionizing continuum.

\begin{itemize}

\item[(i)]{A temporary obscuration of the ionizing continuum source
  incident upon BLR clouds by a moving veil of gas lying between the
  inner accretion disk and the inner BLR.  This could represent an
  episodic ejection of material from the accretion disk.  This veiling
  gas should be moderately highly ionized and of modest column density
  to produce the phenomena described above.    These
  details are left for future investigation.}

\item[(ii)]{A temporary change in the intrinsic ionizing SED
leading to a reduction in the number of ionizing photons,
particularly above $\sim 54$~eV. NGC\,5548 apparently has a hard
ionizing continuum, in which Comptonization plays an important role
(Magdziarz et~al.\ 1998; Mehdipour et al. 2015). 
A temporary
restructuring of the Comptonizing atmosphere situated above the inner
accretion disk could significantly reduce the production of
EUV continuum photons.}
\end{itemize}

Since all of the broad emission lines (UV and optical) display a
deficit in flux during the anomaly, then if temporary obscuration is
the mechanism responsible, this imposes constraints on the nature of
the obscuring veil of material and/or the BLR geometry.  For example,
for the obscuring veil of material to affect all lines simultaneously,
it must cover a large solid angle, thus constraining its
location to be close to the continuum source. Alternatively, if the
BLR is in a flattened configuration, the obscuring veil of
material need not be raised to very large scale heights above the disk
in order to affect all of the emission lines simultaneously, though
such configurations for the BLR may be precluded on energetic
grounds. Indeed, it may even be possible to distinguish between these
differing scenarios by investigating the variation in the start and
end times of the anomalous behavior among different lines. 

Evidence for a change in the disk geometry can be found in high-resolution 
{\it Chandra\/} and {\it XMM-Newton\/} X-ray spectra of
NGC~5548 taken during 2013--2014 (Di Gesu et~al. 2015; Medhipour et
al. 2016).  Based on an observed reduction in the covering fraction of
the X-ray obscurer at higher source luminosities, Di Gesu et
al. (2015) and Medhipour et al. (2016) argue that the continuum-emitting 
region producing the soft excess becomes larger as the source
brightens. If this brightening coincides with a puffing up of a disk
whose outer regions are optically thick at photon energies $E_{\rm ph}
>$\, 54~eV, then this gas may provide the shielding necessary to
explain the observations.

However, while obscuration can plausibly explain the significant drop
in flux and EW of the broad emission lines during the anomaly, it does
not readily explain their reduced amplitude of variability. While an
obscuring screen of material will reduce the flux of ionizing
continuum photons incident upon BLR clouds, the fractional variability
amplitude of the continuum will remain unchanged.  By contrast,
physical changes in the Comptonizing atmosphere altering the SED at
especially EUV photon energies might also lead naturally to
suppressed variability at these energies. The second scenario can
thereby explain not only the significant emission-line flux deficits,
but also the absence of significant emission-line variations
($\eta_{\rm eff} \approx 0$) during the anomalous period. Note that
our emission-line reconstructions (Figures 2a--2d) indicate that if
the ionizing continuum had varied as the 1157\,\AA\ continuum did, we
would have easily detected the corresponding emission-line variations.
In addition, the first scenario might be expected to be accompanied by
a significant increase in the X-ray HR, with greater
photoabsorption occurring for $E_{\rm ph} < 1$ keV, while
observations indicate that for the majority of the anomalous period
the HR remains approximately constant (Figure~2g). On the other
hand, a reorganization of the Comptonizing atmosphere could manifest
itself differently.  Speculating further, the decline in the
variability amplitude of the 1157\,\AA\ continuum during days
$\sim75$--120 resembles a damped oscillation, and could be a
manifestation of a redistribution of energy within the Comptonizing
atmosphere.  A more detailed analysis of the \HST/COS\, data and
contemporaneous X-ray spectroscopy from {\it Chandra\/} and {\it
  Swift\/} (Mathur et~al.\ 2016, in prep.) are currently
underway. When combined, such data could potentially distinguish
these (and possibly other) scenarios, thereby leading to a deeper
understanding of the inner workings of AGN accretion disks.

From the outset of AGN STORM, we had assumed that correlated continuum
and emission-line variations were the primary (and perhaps only) means
of probing the spatial distribution and kinematics of the BLR and the
nature of the central engine. Instead, by good fortune, the presence
of the anomaly, far from compromising the current RM campaign, has
provided a different and rather unique window into the behavior of the
central engine and the processes that drive the continuum and BEL 
variations.

\section{Summary}
\label{section:summary}

Analysis of the time-dependent behavior of the BEL
reprocessing efficiency has proven to be a powerful tool for investigating
the relationship between the ionizing continuum source and line-emitting 
gas. Here we have used it to reveal the origin of the
anomalous behavior exhibited by the strong, broad UV emission
lines during the middle of the 2014 \HST/COS\, monitoring campaign of
the nearby Seyfert 1 galaxy NGC\,5548: a significant yet temporary softening
of the ionizing continuum incident upon BLR clouds.  By following the
energy, we are able to recover the form for the expected
emission-line light curve during the anomaly, as though the anomaly
had not occurred. In so doing, we identified a potential route
for utilizing all of the available data from this campaign in future
attempts at recovering the emission-line response function for the
strongest broad UV emission lines.  

Finally, we note that effects of this nature have been looked for, but
not found, in
previous UV spectroscopic monitoring campaigns of NGC~5548. 
At the level at which these effects are present in the
2014 \HST\, campaign, they would have easily been detected if present,
in both the 1989 and 1993 UV monitoring campaigns on NGC~5548. Nor do
we find such behavior in the far more extensive 13~yr optical
monitoring data available for NGC~5548 (Grier \& Peterson 2003; Goad,
Korista, \& Knigge 2004).  Thus, anomalous behavior of this kind is
likely rare in this source. Since analyses similar to that presented
here have yet to be performed on the archival UV and optical datasets
available for other AGN, we are unable to comment on the occurrence
rate of this phenomenon among the RM AGN.

\begin{figure*}
\includegraphics[angle=0,scale=0.80]{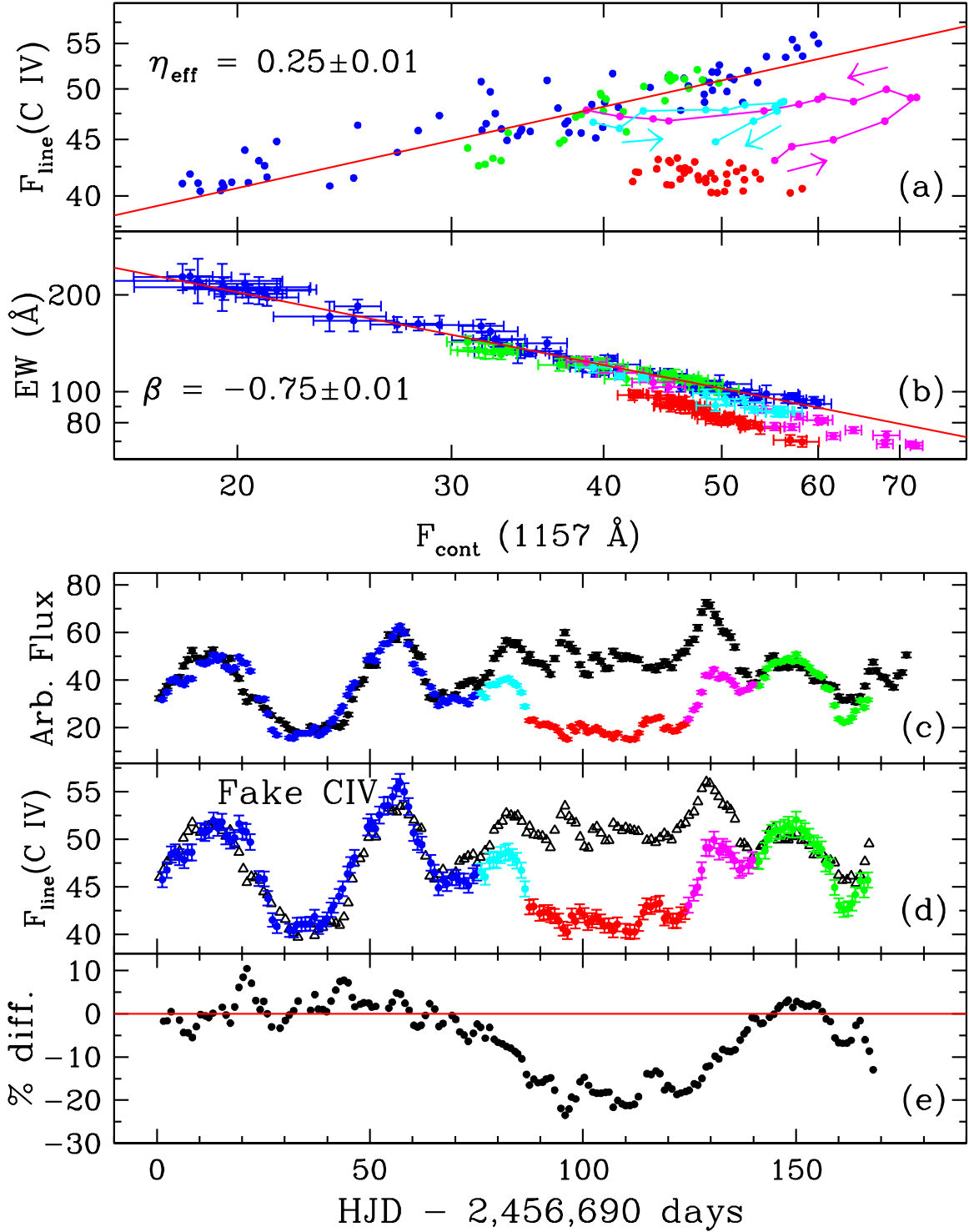}
\caption{Upper two panels -- (a) $\log F$(C~{\sc iv}) versus
  $\log F_{\rm cont}$(1157\,\AA), color coded as follows:
  HJD$-$2,400,000: 56,691.2813--56,765.1133 (blue),
  56,766.1094--56,776.3984 (cyan), 56,777.4336--56,813.9688 (red),
  56,814.7891--56,829.8281 (magenta), 56,830.8242--56,856.8438
  (green). Error bars have been omitted for clarity. In particular,
  note the island of red points lying away from the main relation, as
  well as the cyan and magenta points which trace a contiguous path to
  and from the main relation. The red line indicates the best-fit
  slope to the blue and green points only. (b) log EW(C~{\sc iv})
  versus log $F_{\rm cont}$(1157\,\AA). As above, the red line
  indicates the best-fit slope to the blue and green points
  only. Bottom three panels -- (c) A comparison between the scaled and
  shifted version of the C~{\sc iv} emission-line light curve (colored
  points) and the observed 1157\,\AA\ continuum light curve
  (black points), highlighting the anomalous flux and response in this
  emission line midway through the campaign.  (d) A comparison
  between the observed C~{\sc iv} emission-line light curve (colored
  points) and the reconstructed emission-line light curve (black
  triangles; see text for details). (e) The estimated percentage
  deficit in the C~{\sc iv} flux.}
\label{fig1}
\end{figure*}

\begin{figure*}
\includegraphics[angle=0,scale=0.8]{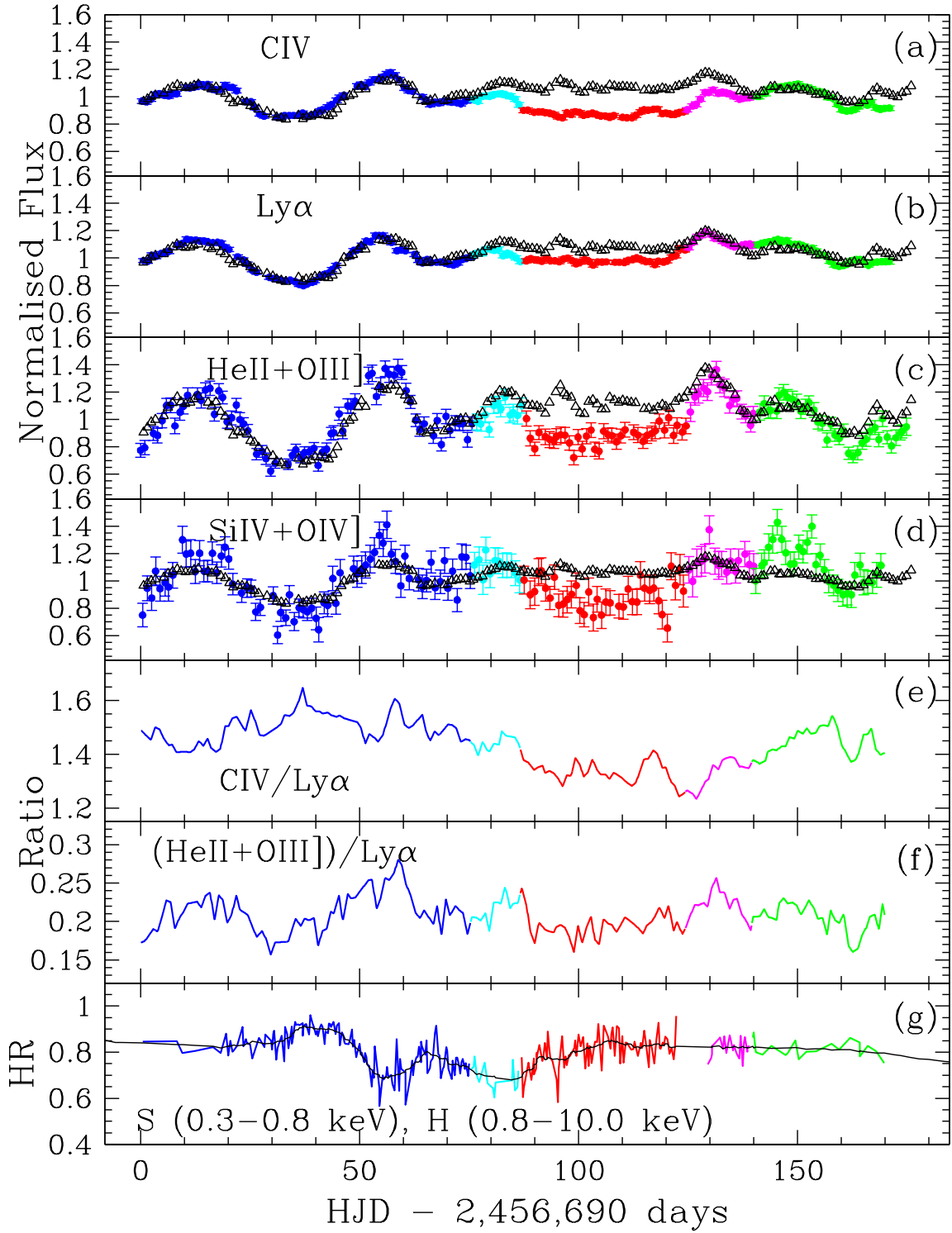}
\caption{Panels (a)--(d), a comparison between the observed
  BEL light curves (color coded as described in the text)
  and their respective {\it reconstructed\/} light curves (black
  triangles). Each has been shifted backward in time according to the
  measured delay between the 1157\,\AA\ continuum and emission-line light
  curve for that particular line, determined over the first 75 days.
  Panels (e)--(f), time-variable emission-line ratios C~{\sc
    iv}/Ly$\alpha$ and (He~{\sc ii}$+$O~{\sc iii]})/Ly$\alpha$. (g)
The time-variable {\it Swift}/XRT X-ray hardness ratio (H$-$S/H$+$S)
color coded as above). The solid black line indicates the hardness
ratio after smoothing with a 15-point boxcar filter.}
\label{fig2}
\end{figure*}

\clearpage

\begin{table}
\begin{center}\caption{BEL delays and adopted narrow emission-line fluxes\label{tab1}}
\begin{tabular}{cccc}
\tableline 
\tableline
Line ID & CCF(cent) & CCF(lag) & F(narrow)\\
         & (days) & (days) & ($10^{-13}$~erg~s$^{-1}$~cm$^{-2}$)\\
\tableline
Ly$\alpha$                  & 6.69 $\pm$ 0.41 & 6.41 $\pm$ 0.51 & 8.9 \\
Si~{\sc iv} $+$ O~{\sc iv}] & 5.80 $\pm$ 0.58 & 5.86 $\pm$ 0.89 & 1.2 \\
C~{\sc iv}                  & 4.97 $\pm$ 0.39 & 4.95 $\pm$ 0.50 & 7.0 \\
He~{\sc ii}$+$ OIII]  & 2.42 $\pm$ 0.44 & 2.14 $\pm$ 0.58 & 1.2 \\
\tableline
\end{tabular}

\end{center}
\end{table}

\clearpage

\begin{table}
\begin{center}\caption{Time-averaged BEL responsivities and the slope of the intrinsic Baldwin effect\label{tab2}}
\begin{tabular}{ccc}
\tableline
\tableline
Line ID & $\eta_{\rm eff}$ & $\beta$ \\
\tableline
Ly$\alpha$                  & 0.30 $\pm$ 0.01 & $-$0.73 $\pm$ 0.02 \\
Si~{\sc iv} $+$ O~{\sc iv}] & 0.45 $\pm$ 0.01 & $-0$.58 $\pm$ 0.03 \\
C~{\sc iv}                  & 0.25 $\pm$ 0.01 & $-$0.75 $\pm$ 0.01 \\
He~{\sc ii} $+$ OIII]       & 0.58 $\pm$ 0.04 & $-$0.48 $\pm$ 0.04 \\
\tableline
\end{tabular}

\end{center}
\end{table}

\begin{table}
\begin{center}\caption{Time-averaged BEL EWs for the anomaly and the main relation\label{tab3}}
\begin{tabular}{crr}
\tableline
\tableline
Line ID & EW (\AA)  & EW (\AA) \\
        & (anomaly) & (main relation) \\
\tableline
Ly$\alpha$                  & 64.5 $\pm$ 5.1 & 72.5$\pm$ 5.9 \\
Si~{\sc iv} $+$ O~{\sc iv}] & 6.2 $\pm$ 0.9  & 8.2$\pm$ 1.2 \\
C~{\sc iv}                  & 86.2 $\pm$ 6.9 & 106.3$\pm$ 8.3 \\
He~{\sc ii} $+$ OIII]       & 12.5 $\pm$ 1.3 & 15.9$\pm$ 3.3 \\
\tableline
\end{tabular}

\tablenotetext{}{The EWs are here measured with respect to the
  average UV continuum flux at 1157\,\AA\ for those epochs
  spanning the time of the anomaly (Figure~1a -- red points); $F_{\rm cont}$(1157\,\AA) $= (4.84\pm0.37) \times 10^{-14}$~erg~s$^{-1}$~cm$^{-2}$~\AA$^{-1}$.}

\end{center}
\end{table}

\acknowledgments

{\it Facilities:} \facility{\HST (COS)}.




\acknowledgments 

Support for \HST\ program number GO-13330 was provided by NASA through
a grant from the Space Telescope Science Institute,
which is operated by the Association of Universities for Research in
Astronomy, Inc., under NASA contract NAS5-26555.
M.M.F., G.D.R., B.M.P., C.J.G., and R.W.P.\ are grateful for the support of the
National Science Foundation (NSF) through grant AST-1008882 to The Ohio
State University. 
A.J.B.\ and L.P.\ have been supported by NSF grant AST-1412693. 
A.V.F.\ and W.-K.Z. are grateful for financial assistance from NSF grant 
AST-1211916, the TABASGO Foundation, and the Christopher R. Redlich Fund.
M.C.\ Bentz gratefully acknowledges support through NSF CAREER grant AST-1253702 to Georgia State University.
M.C.\ Bottorff acknowledges HHMI for support through an undergraduate science education grant to
Southwestern University.
K.D.D.\ is supported by an NSF Fellowship awarded under grant AST-1302093.
R.E.\ gratefully acknowledges support from NASA under awards NNX13AC26G, NNX13AC63G, and NNX13AE99G.
J.M.G.\ gratefully acknowledges support from NASA under award NNH13CH61C.
P.B.H.\ is supported by NSERC. 
K.D.H.\ acknowledges support from the UK Science and Technology Facilities Council
through grant ST/J001651/1.
M.I.\ acknowledges support from the Creative Initiative program, No. 2008-0060544, of the National Research Foundation of Korea (NRFK) funded by the Korean government (MSIP).
M.D.J.\ acknowledges NSF grant AST–0618209.
SRON is financially supported by NWO, the Netherlands Organization for Scientific Research.
B.C.K.\ is partially supported by the UC Center for Galaxy Evolution.
C.S.K.\ acknowledges the support of NSF grant AST-1009756.  
D.C.L.\ acknowledges support from NSF grants AST-1009571 and AST-1210311.
P.L.\ acknowledges support from Fondecyt grant \#1120328.
A.P.\ acknowledges support from an NSF graduate fellowship and a UCSB Dean's
Fellowship. 
J.S.S.\ acknowledges CNPq, National Council for Scientific and Technological Development (Brazil)
for partial support and The Ohio State University for warm hospitality.
T.T.\ has been supported by NSF grant AST-1412315.
T.T.\ and B.C.K.\ acknowledge support from the Packard Foundation in the form of a Packard Research
Fellowship to T.T; also, T.T.\ thanks the American Academy in Rome and the Observatory
of Monteporzio Catone for kind hospitality.
The Dark Cosmology Centre is funded by the Danish National Research Foundation.
M.V.\ gratefully acknowledges support from the Danish Council for Independent Research via grant no. DFF – 4002-00275.
J.-H.W.\ acknowledges support by the National Research Foundation of Korea (NRF) grant funded by the Korean government (No. 2010-0027910).
This research has made use of
the NASA/IPAC Extragalactic Database (NED), which is operated by the
Jet Propulsion Laboratory, California Institute of Technology, under
contract with NASA.

\end{document}